\documentstyle[12pt,epsf]{article}
\title{Delta excitation in antiproton-deutron annihilation
\thanks{Supported by the BMBF.}}
\author{A. Sibirtsev\thanks{Permanent address:
                 Laboratory of Nuclear Problems, Joint Institute of
                 Nuclear Research, Dubna, 141980, Russia
        and      Institute of Theoretical and Experimental Physics,
                 Moscow, Russia.},
K. Tsushima\thanks{Address since October:
Department of Physics and Mathematical Physics, The University
of Adelaide, Adelaide 5005 Australia.}
and Amand Faessler\\
Institut f\"ur Theoretische Physik, Universit\"at T\"ubingen,\\
Auf der Morgenstelle 14, D-72076 T\"ubingen, Germany}
\begin{document}
\maketitle
\begin{abstract}
The $\Delta$-excitation in $\overline{p}d$ annihilation at
rest was studied. The annihilation amplitude from the
statistical model and the $\pi N$ amplitude
from the resonance model were adopted in our calculations.
We analyze the invariant mass
of the $\pi^+p$ and $\pi^-p$ systems selecting the protons
with momenta above 400 MeV/c and with respect to the
different final reaction channels. Our model reproduces reasonably the
experimental data.

\end{abstract}
\section{Introduction}
One of the unique features of the annihilation of stopped antiprotons
in nuclei is the production of very fast moving
baryons. It was suggested
by Armenteros and French~\cite{Armenteros} that the high momentum
tail of the proton spectra in antiproton annihilation may
be ascribed to the annihilation involving more than one nucleon.
As was found by
Hernandes and Oset~\cite{Hernandez1} that a sizable increase of the
fast moving nucleons are produced by the
many-body annihilation mechanism.
Furthermore, Cugnon et al.~\cite{Cugnon} suggested
that the production of fast moving
strange baryons implies the two-body annihilation mechanism.

However, a quite different mechanism of
fast  protons formation
was suggested by Kudryavtsev and
Tarasov~\cite{Kudryavtsev}.They suggest  that the
fast protons are produced in
antiproton-deuteron annihilations  due to
pion absorption on the recoiling nucleon. Moreover the protons with
very high momenta may be produced from the absorption and the
emission of a meson by a baryon resonance.

One of the traditional mechanisms of fast baryon production
following the antiproton annihilation is the secondary interactions of the
pions as well as the mesonic resonances in the nuclear environment.
It was shown by Sibirtsev~\cite{Sibirtsev1} that the high momentum tail of
the proton spectra from carbon and uranium targets
can be reproduced with the rescattering and
absorption of the annihilation mesons.
Fasano et al.~\cite{Fasano} found that the pion rescattering
can explain the fast proton production in the reaction $\overline{p}d
\rightarrow 5\pi p$ but strongly underestimates the proton spectrum from
reaction $\overline{p}d \rightarrow 3\pi p$.

It is clear that an
essential indication for the contribution of the meson rescattering
is the excitation of the $\Delta$-resonances.
Obviously the direct experimental evidence of secondary interactions
of the mesons produced from annihilation is the observation
of the isobar structure in the effective mass of the $\pi N$-system.

The search for isobar was performed in $\overline{p}d$ annihilation
by Kalogeropoulos et al.~\cite{Kalogeropoulos}, but they found no
$\Delta$ influence in   $\pi N$-system. However as was later shown
by Voronov and Kolybasov~\cite{Voronov} analyzing the reaction
$\overline{p}d \rightarrow 2\pi^+3\pi^-p$ that the structure of
the isobar may be significantly smeared out due to  the
contribution from  $\pi N$-pairs in which the pion does not undergo
rescattering. Moreover it was suggested that the $\Delta$-resonance
can be observed only when the recoil nucleons are selected with
momenta above 200 MeV/c.

Recently the  OBELIX collaboration from CERN measured the
annihilation of stopped antiprotons in a deutron target~\cite{Ableev}.
The invariant  mass of protons with momenta larger than 400 MeV/c
and $\pi^+$-mesons indicates a clean peak from $\Delta^{++}$-resonances,
whereas the $\Delta^0$-resonance in the $\pi^-p$ system was not found.
Presently the isobar production in antiproton-deuteron annihilation is
measured with respect to the different final
channels~\cite{Private}.

In the present paper we study the $\Delta$-excitation in
$\overline{p}d$ annihilation at rest. The dependence of the
$\Delta^{++}$-production on the pion multiplicity is analyzed and the
invariant mass of the $\pi N$-system for the  reaction
channels are predicted.

\section{The model}
Similar to~\cite{Kolybasov1} we
calculate the amplitude of the triangle diagram shown in fig.1 as
\begin{equation}
T = \frac {E_{\pi}+E_p} {2 \pi (m-E_{\pi})}
\int \frac {T_1(\overline{p}N \rightarrow n\pi) T_2(\pi N)}
{{\bf k}_1^2 - {\bf k}_2^2+i\epsilon} \phi ({\bf k}_1+{\bf Q}/2)
d{\bf k}_1
\end{equation}
where $T_1$ is the amplitude of $\overline{p}N \rightarrow n \pi$
annihilation, $E_{\pi}$ and $E_p$ are the pion and
the proton energies in the final state, ${\bf k}_1$ and ${\bf k}_2$
are the pion momenta before and after the interaction. Here $m$ is the nucleon
mass and ${\bf Q}$ is the deutron momentum.
The deutron wave function $\phi ({\bf Q})$
 for the Bonn potential~\cite{Bonn1} was adopted.

We use the annihilation probability from~\cite{Hernandez} as
\begin{equation}
\label{am1}
| T_1(\overline{p}N \rightarrow n\pi) | ^2 =
G_f \frac {\lambda^{1/2}(s,
m^2,m^2)}{2m^2} \frac {P_n} {I_n(s)}
\end{equation}
where  $s$ is the squared invariant mass of
$\overline{p}N$ system and $\lambda$ is the K\"alen function.
Factor $G_f$ stands for the charge configuration of the final system of
$n$-pions and was calculated with the statistical approach as~\cite{Pais}
\begin{equation}
\label{stat}
G_f= \left[n_+! n_-! n_0! \right]^{-1}
 \left[ \sum_{\beta} \left(n_+! n_-! n_0! \right)_{\beta}^{-1}
\right]^{-1}
\end{equation}
where $n_+$, $ n_-$ and $ n_0$ are the numbers of positive, negative and
neutral $\pi$-mesons for a final charge system of $n=n_++n_-+n_0$
pions and the summation is performed over all reaction channels allowed for a
given $n\pi$ system. Let us  note, that the more complicated
expression for $G_f$ suggested in~\cite{Iljinov} is quite similar
to the statistical factor~(\ref{stat}).

In eq.(\ref{am1}) the factor $I_n(s)$ accounts for the phase space volume of
$n$-pions with invariant mass $\sqrt{s}$ and $P_n$ stands for the probability
for the  creation of $n$-pions in $\overline{p}N$ annihilation, which was
taken as~\cite{Stenbacka}
\begin{eqnarray}
P_n= \left(2\pi \sigma \right)^{1/2} exp \left[ -\frac {(n-\nu)^2}
{2\sigma} \right] \nonumber \\
\nu =2.65+1.78 lns, \ \ \ \sigma=0.174 \nu s^{0.2}
\end{eqnarray}
which are different from $\nu=5.05$ and $\sigma =0.76$
suggested in~\cite{Iljinov} for the annihilation at rest, but as we
studied~\cite{Sibirtsev2} do not change the results noticably.

Accordingly to Hernandez et al.~\cite{Hernandez} the annihilation amplitude
has a smooth momentum dependence and the momentum distribution of the
pions are described by the $n$-body phase space.

The $T_2(\pi N )$ amplitude accounts for the
$\pi N \rightarrow \pi N$ scattering in the $\Delta$-isobar region
and was calculated with the resonance model~\cite{Tsushima1,Tsushima2}.
The cut-off parameter $\Lambda$ was fitted to the experimental data of the
total cross section for the $\pi^+p$ interaction~\cite{Landolt}
(see fig.2). The lines show our calculations
with the resonance model and the
parameters $ \Lambda$=0.3 GeV solid,
0.35  GeV dashed, 0.4  GeV dotted and 0.5 GeV dashed-dotted line.
In the futher calculations we use $\Lambda$=0.33 GeV.

With fig.3 we demonstrate how reasonable the resonance model
reproduces the angular spectra of the $\pi$-mesons. Here the dots show
the experimental data on the differential cross sections of the
reaction $\pi^+ p \rightarrow \pi^+ p$ plotted as a function
of the $cos\theta$ in the c.m.s. system. The squares show the
results for $s=1.251$ GeV$^2$~\cite{Auld}, the triangles
for  1.474~\cite{Bussey} and the circles for 1.711~\cite{Gordeev} .
The solid, dashed and dotted lines show our results calculated with
the resonance model. The dashed-dotted line shows the function
\begin{equation}
\label{fcos}
\frac {d\sigma} {d\Omega} \propto 1+3cos^2\theta
\end{equation}
which is  valid in the central $\Delta$-resonance region.
Our results obtained with the resonance model are also in
reasonable agreement with the Legendre expansions of
the differential cross sections calculated from the
Karlsruhe-Helsinki partial wave solution.

\section{Results}
The invariant mass distribution of the $\pi^+p$ system
produced  in the  annihilation of antiprotons on the deuteron
at rest in the reaction
$\overline{p}d \rightarrow p \pi^+ 2\pi^- m\pi^0$ are shown in
Fig.4. The experimental data were taken from~\cite{Ableev}.
Similar as in ref.~\cite{Ableev} we select the protons with momenta
above 400 MeV/c. The dashed line shows the $\Delta^{++}$-resonance,
whereas the dotted line shows the combinatorial backround that comes
from the pions which do not undergo interactions. The solid line is the
sum and describes  reasonably the experimental data. Note that
the isobar resonance is slightly shifted to a
higher mass in our calculations, which is due to the
selection of the fast protons with
momenta above 400 MeV/c. Surprisingly this obvious feature was
not observed in the experimental study~\cite{Ableev}.
The isobar structure is very clearly  detected and reconstructed
from the total invariant mass spectrum.

In fig.5 we show the contribution from the isobar (a) and the
bakground (b) calculated for the reaction
$\overline{p}d \rightarrow p \pi^+ 2\pi^- m\pi^0$  for
$m=0$ dotted, for $m=1$ dashed, for $m=2$ solid and
for $m=3$ dashed-dotted line.
It is clear that the dominant contribution comes from the reaction
channels with one and two neutral pions in the final state.
The contribution to the $\Delta$-excitation in the reaction
$\overline{p}d \rightarrow p \pi^+ 2\pi^-$ is negligible, because
the pions created in the annihilation have  momenta which lead
with a nucleon the excitations above the $\Delta$ resonance region.

A quite different situation exists for the $\pi^-p$ invariant mass spectrum
from the reaction $\overline{p}d \rightarrow p \pi^+ 2\pi^- m\pi^0$,
which is shown in Fig.6.
The signal from the $\Delta^0$ resonance is to weak to be extracted from
strong combinatorial background.

With our model we can also  predict the
isobar structure in the reaction channels which are now under study
at CERN.
Fig.7a shows the $\pi^+p$ invariant mass spectrum from the
reaction $\overline{p}d \rightarrow p 2\pi^+ 3\pi^-$. Again we
select  protons with momenta above 400 MeV/c.
Although the contribution from $\Delta^{++}$ is strong, it might be
a difficult  problem to extract it from the total spectrum,
shown as the solid line.
In fig.7b we show the $\pi^+p$ invariant mass spectrum  from the reaction
$\overline{p}d \rightarrow p 2\pi^+ 3\pi^- \pi^0$. The isobar
structure can be clearly seen.

\section{Conclusion}
The $\Delta$-excitation in $\overline{p}d$ annihilation at
rest was studied. We use the annihilation amplitude from the
statistical model~\cite{Hernandez}. The $\pi N$ amplitude
was calculated with the resonance model accounting only for the
$\Delta (1232)$ resonance. We calculate the invariant mass
of the $\pi^+p$ and $\pi^-p$ systems selecting the fast protons
with momenta above 400 MeV/c.

We reasonably reproduce the new experimental
data from the OBELIX collaboration at CERN.
We found that the $\Delta^0$ structure can not be observed in the
$\pi^-p$ invariant mass spectrum because of the strong contribution
from the combinatorial background, in which the pions do not undergo an
interaction.

Most promising  is to study the $\Delta^{++}$ excitation in the
reactions $\overline{p}d \rightarrow p \pi^+ 2\pi^- \pi^0$ and
$\overline{p}d \rightarrow p \pi^+ 2\pi^- 2\pi^0$.
We also predicted that the observation of the $\Delta^{++}$
resonance in the reaction
$\overline{p}d \rightarrow p 2\pi^+ 3\pi^- \pi^0$
is significantly easier than with the reaction
$\overline{p}d \rightarrow p 2\pi^+ 3\pi^-$.

\newpage
\vspace*{2cm}
\noindent{\Large {\bf Figure captions}}\\ \\
\noindent Figure 1: Pion rescattering diagram.

\noindent Figure 2: The total $\pi^+p$ cross section. Experimental data
are from~\cite{Landolt}. Lines show our calculations with
cut-off parameters: 0.3 -solid, 0.35-dashed, 0.4-dotted and
0.5 GeV- dashed-dotted.

\noindent Figure 3: The angular spectra of pions from the reaction
$\pi^+p \rightarrow \pi^+p$ in the c.m.s. system. Experimental
data are from~\cite{Auld,Bussey,Gordeev} for  $s=1.251$ GeV$^2$
-squares, 1.474-triangles and 1.711-circles.
The solid, dashed and dotted lines show our results calculated with
the resonance model. The dashed-dotted line shows~(\ref{fcos}).


\noindent Figure 4: The invariant mass spectrum of the $\pi^+p$
system in the reaction
$\overline{p}d \rightarrow p \pi^+ 2\pi^- m\pi^0$ for
protons with momenta above 400 MeV/c. Experimental data are
from~\cite{Ableev} and lines show our results. Dashed line shows
the contributon from
$\Delta^{++}$-excitation and dotted from the bakground. Solid is the
sum.

\noindent Figure 5: The invariant mass spectrum of the $\pi^+p$
system in the reaction
$\overline{p}d \rightarrow p \pi^+ 2\pi^- m\pi^0$ for
protons with momenta above 400 MeV/c. Fig.a)  shows
the contributon from
$\Delta^{++}$-excitation and Fig.b) from the bakground.
Dotted lines show the contribution from $m=0$, dashed-$m=1$,
solid-$m=2$ and dashed-dotted-$m=3$.

\noindent Figure 6: The invariant mass spectrum of the $\pi^-p$
system in the reaction
$\overline{p}d \rightarrow p \pi^+ 2\pi^- m\pi^0$ for
protons with momenta above 400 MeV/c. Experimental data are
from~\cite{Ableev} and lines show our results. Dashed line shows
the contributon from
$\Delta^{0}$-excitation and dotted from the bakground. Solid is the
sum.

\noindent Figure 7: The invariant mass spectrum of the $\pi^+p$
system in the reaction
$\overline{p}d \rightarrow p 2\pi^+ 3\pi^-$ (a)
and $\overline{p}d \rightarrow p 2\pi^+ 3\pi^- \pi^0$ (b) for
protons with momenta above 400 MeV/c. Dashed lines show
the contributon from
$\Delta^{++}$-excitation and dotted from the bakground.
Solid lines are the sum.
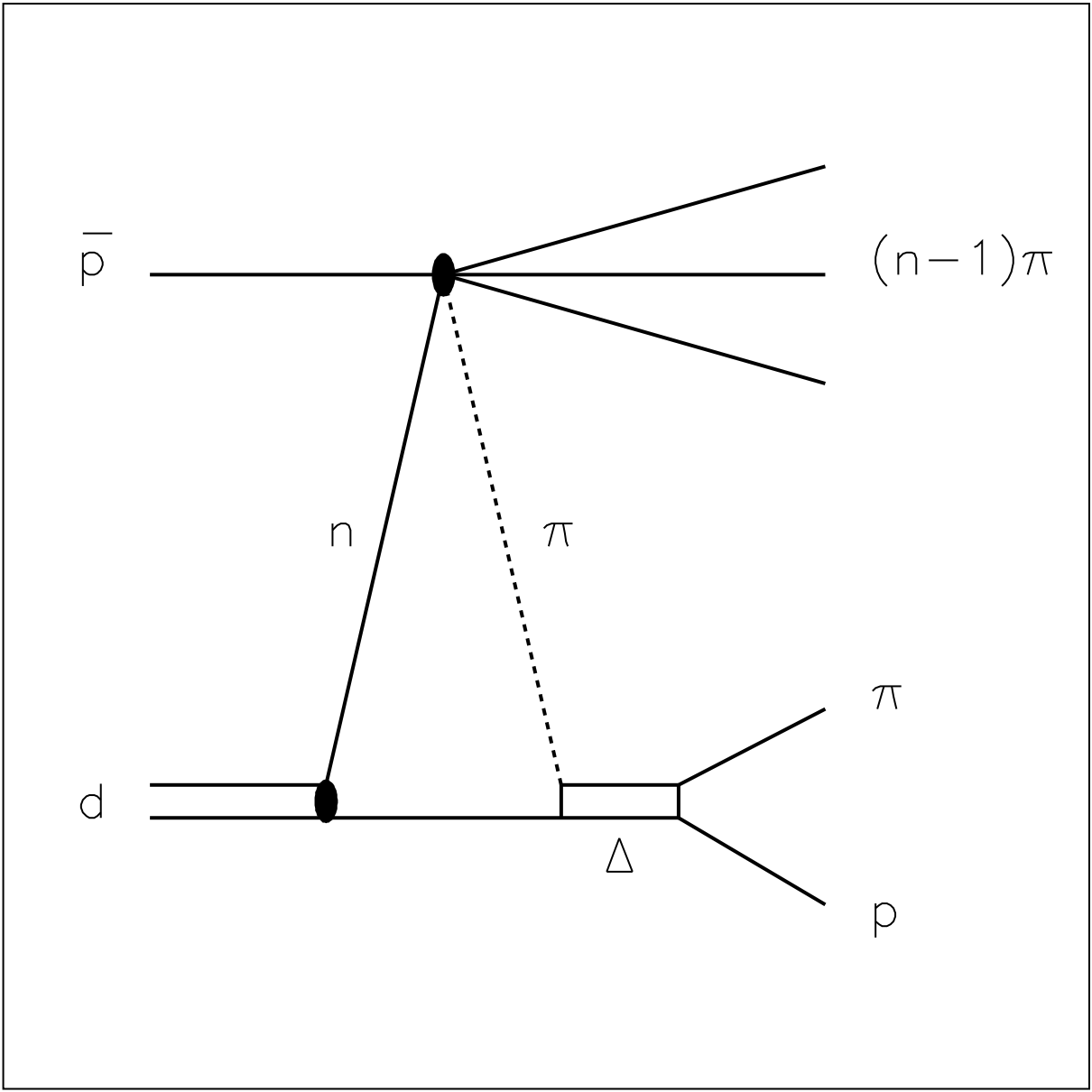
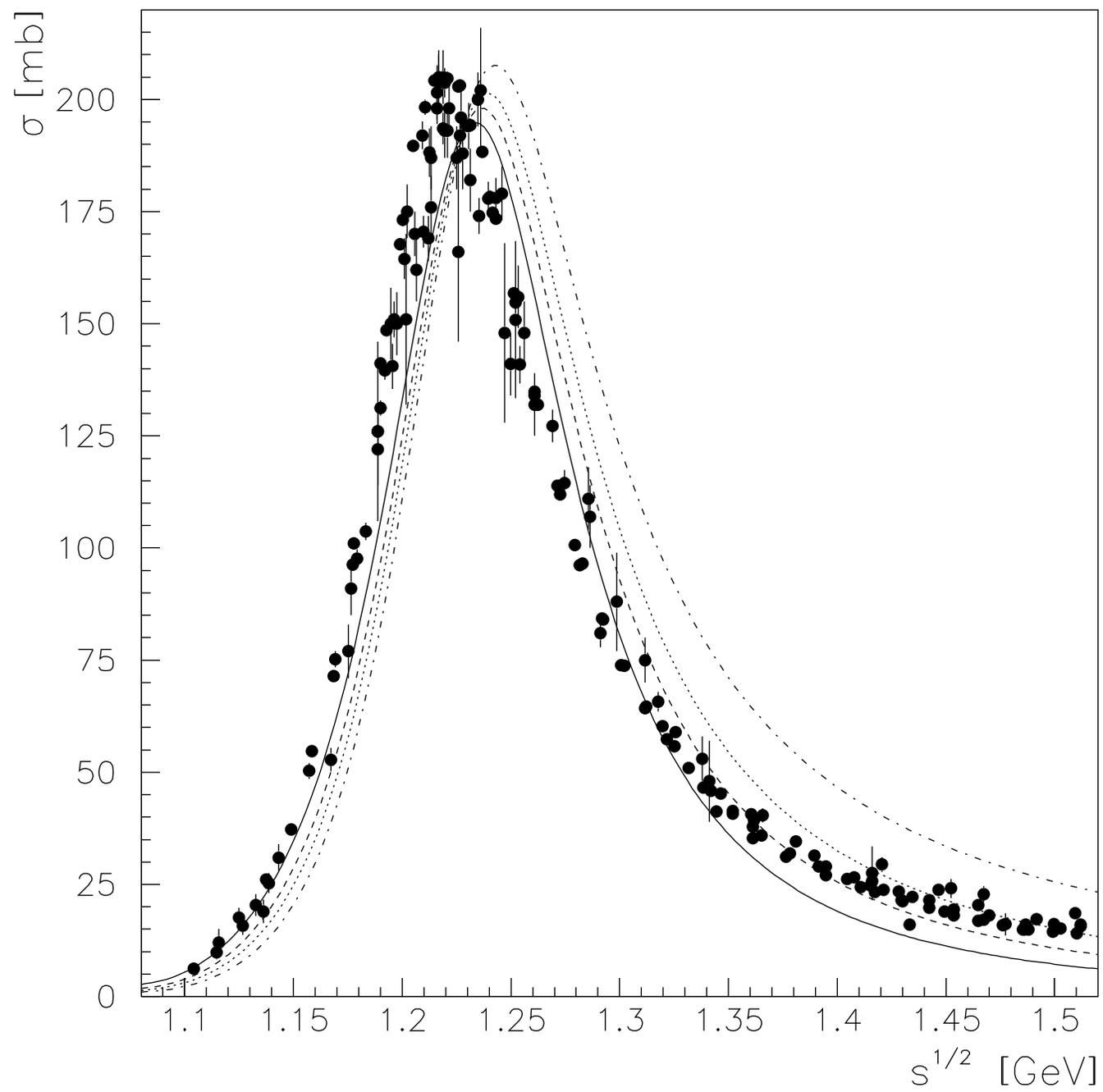
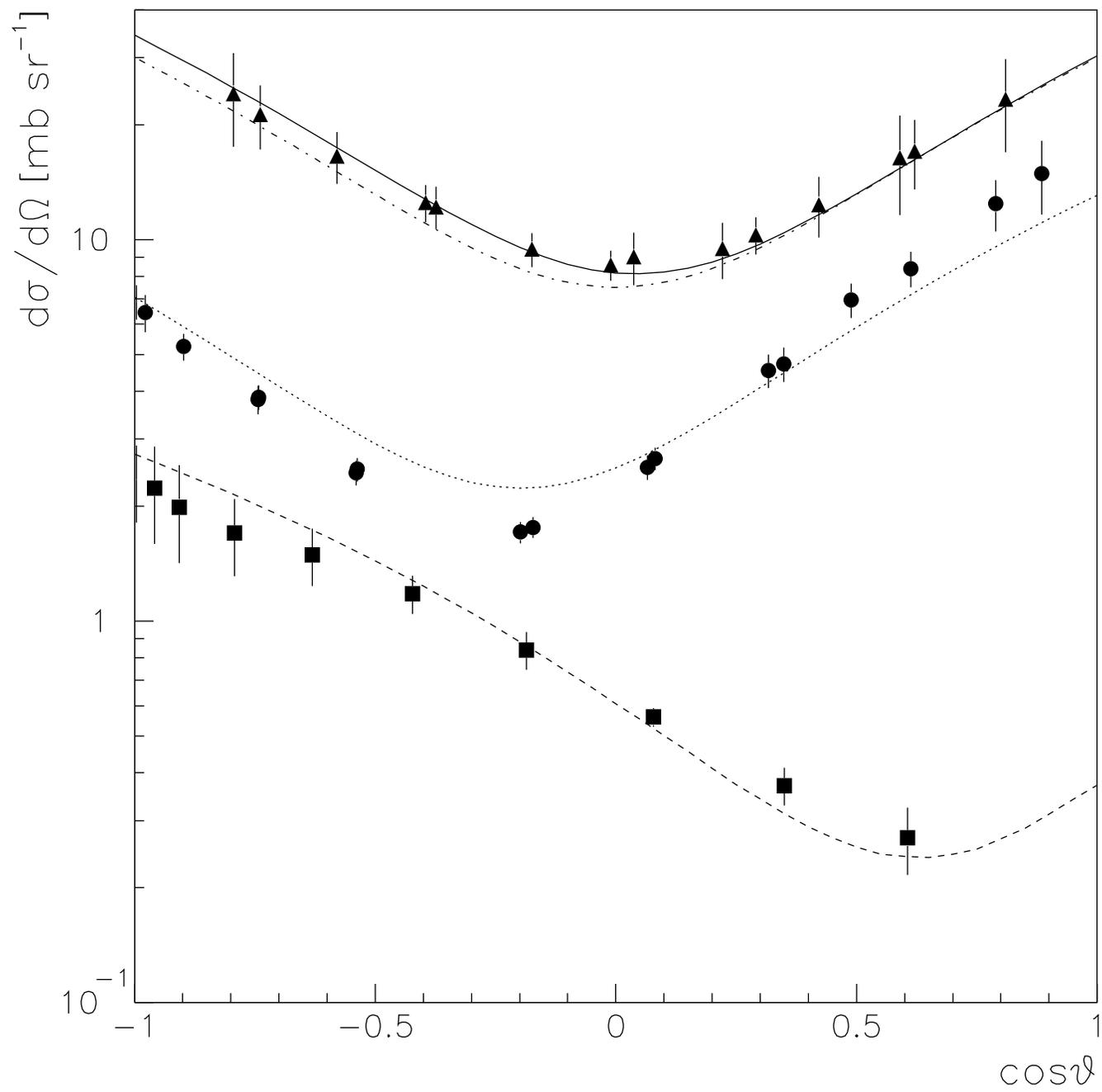
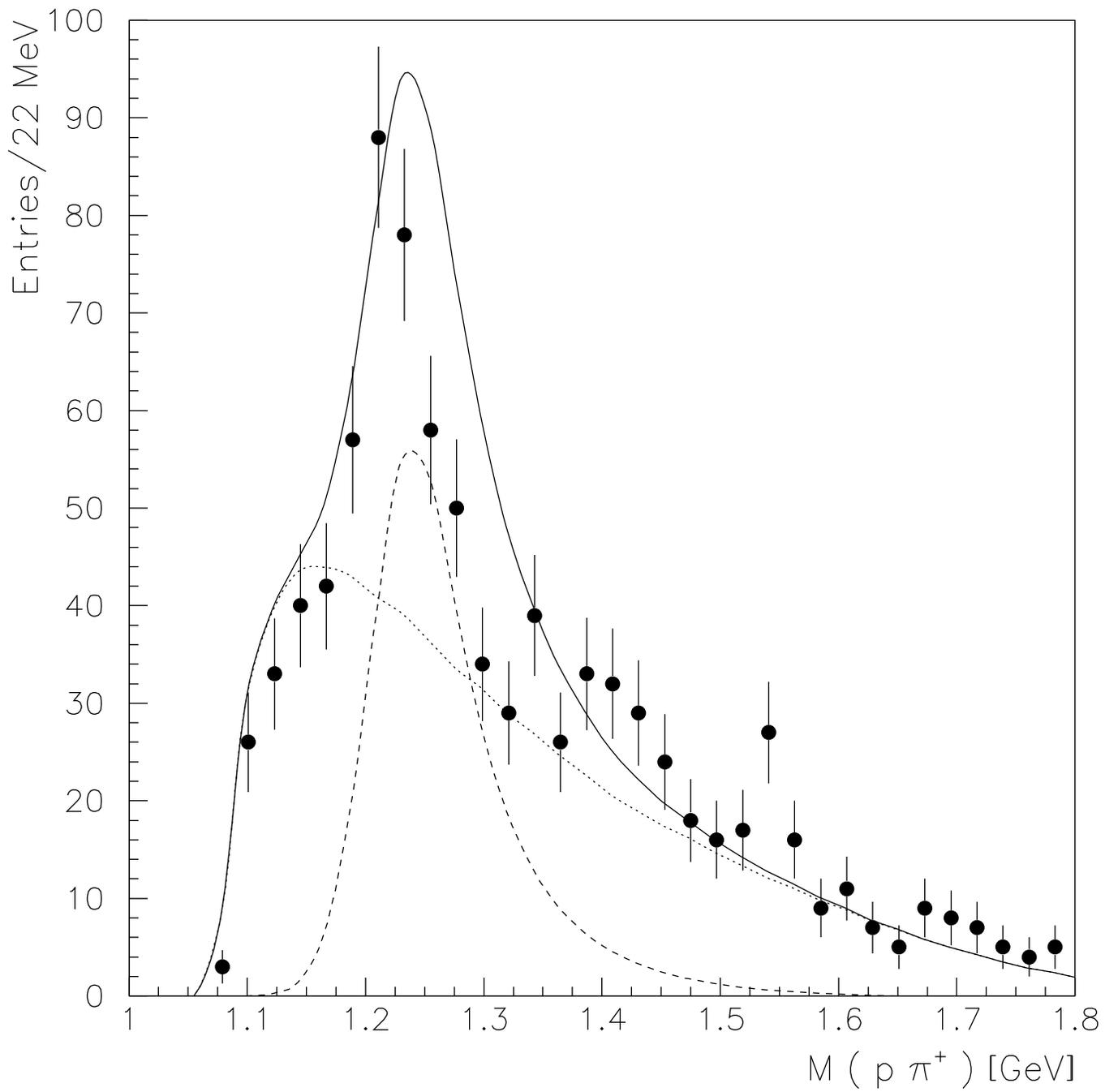
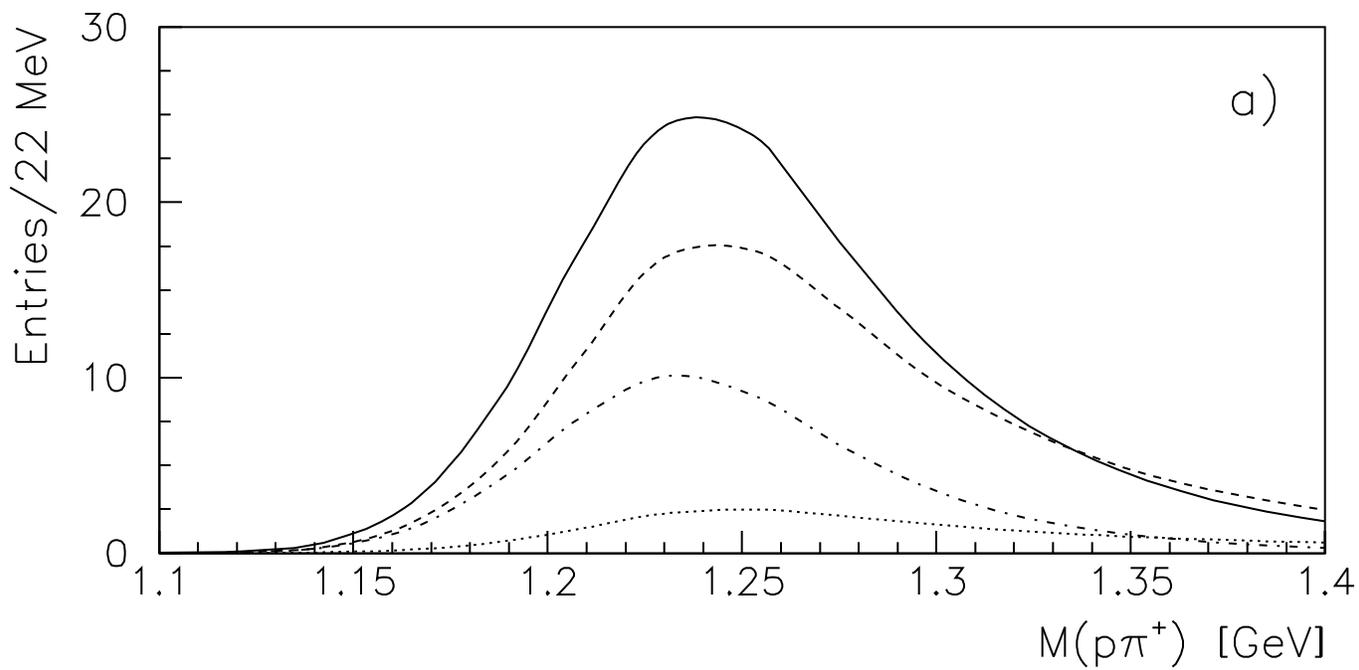
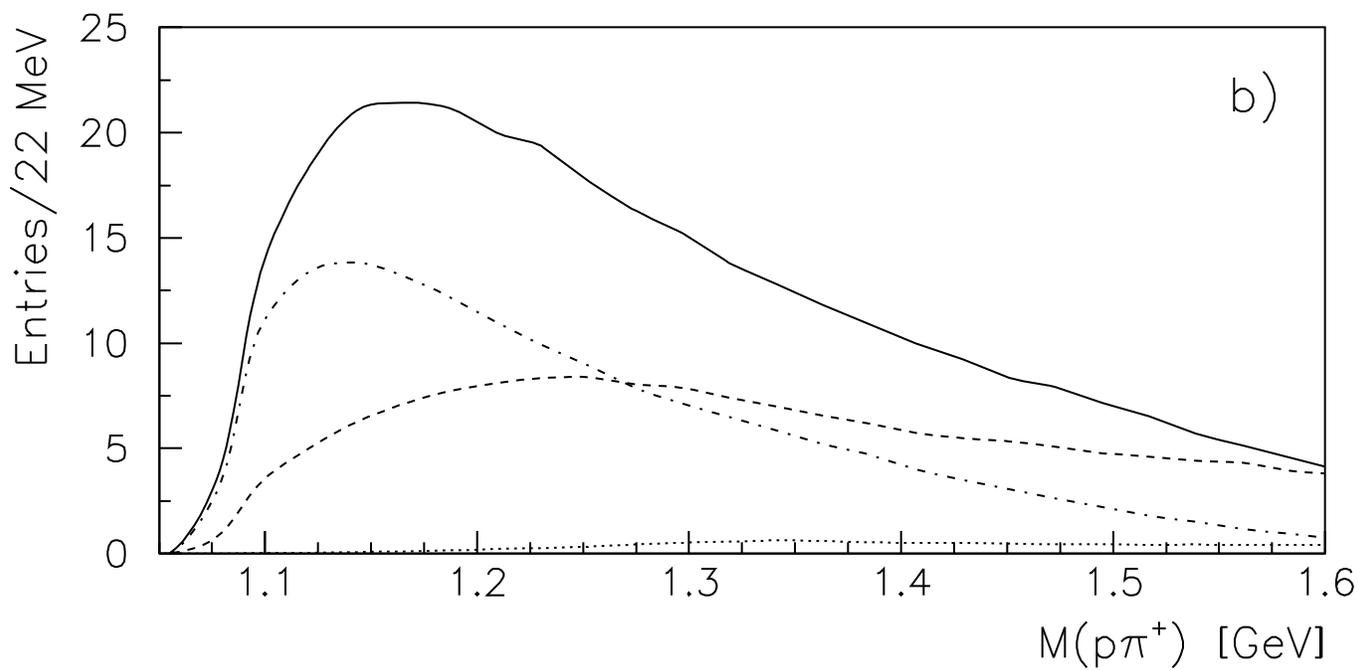
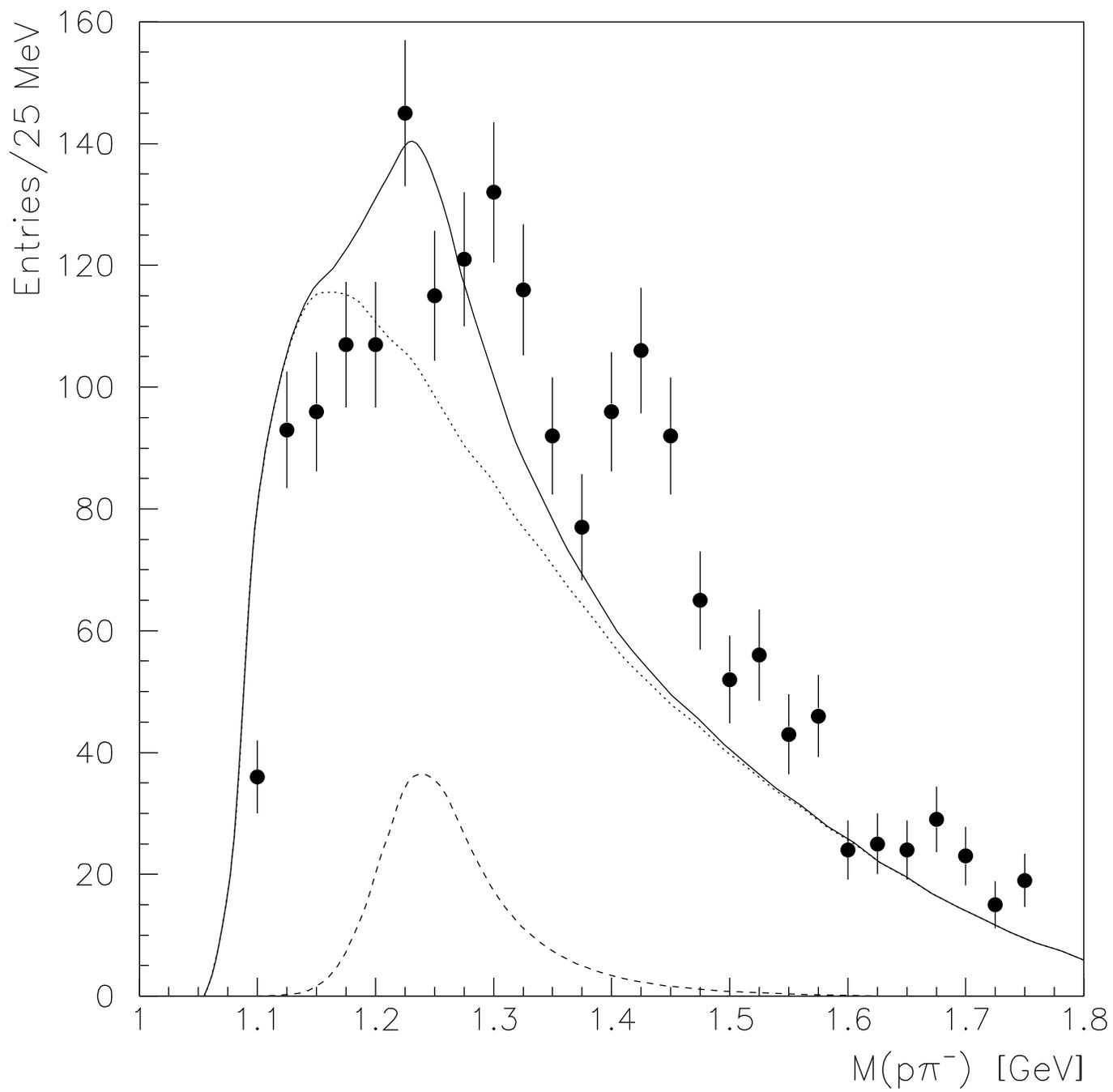
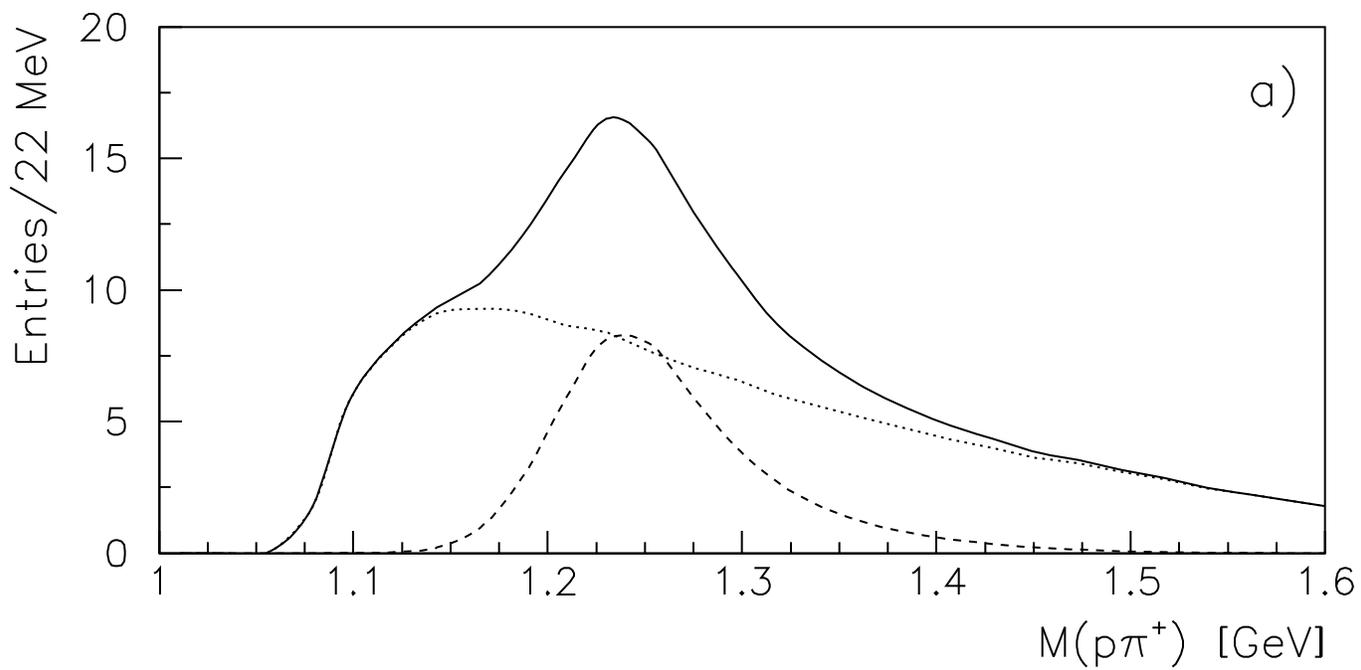
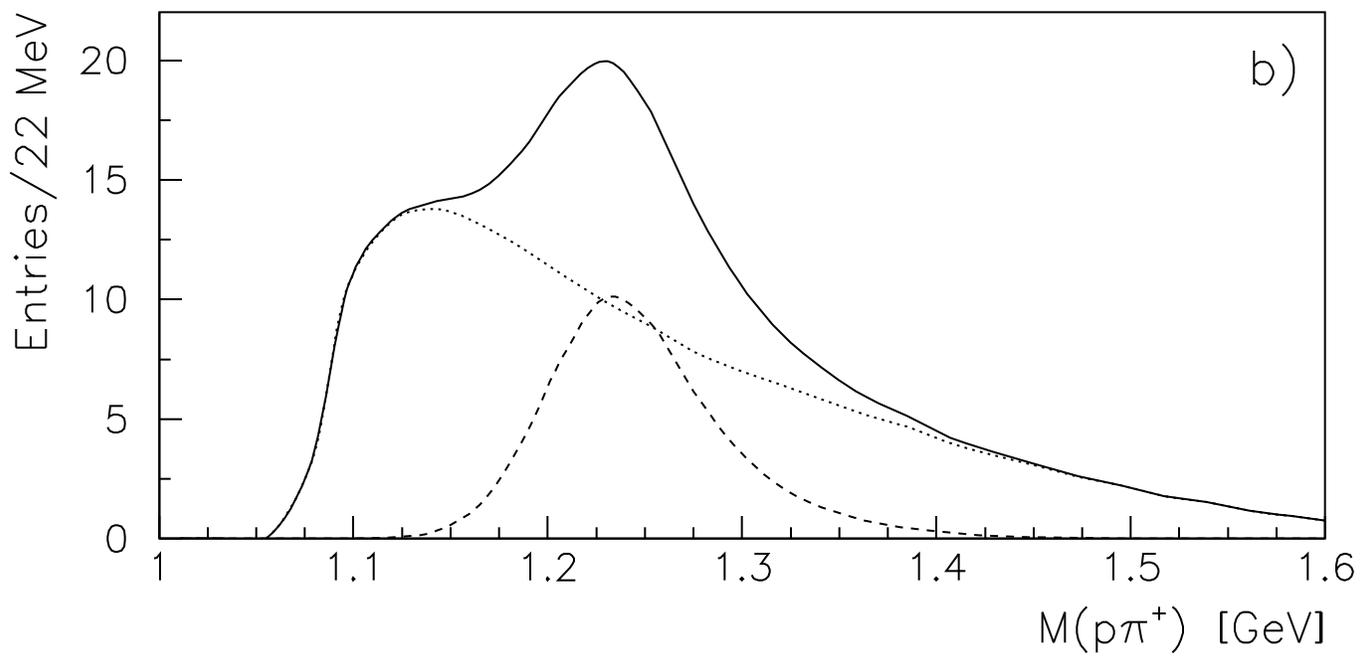
\end{document}